# A review of Plesiochronous Digital Hierarchy (PDH) and Synchronous Digital Hierarchy (SDH)

Olabenjo Babatunde[1], Salim Mbarouk[2]
[1](Department of Information Systems Engineering,
Cyprus International University, Nicosia, Cyprus)
[2] (Department of Computer Engineering,
Cyprus International University, Nicosia, Cyprus)

**ABSTRACT**
With the advancement in telecommunications, packet traffic is rapidly becoming the mainstream of data traffic. The use and deployment of Synchronous Digital Hierarchy (SDH) networks for interconnection has gained traction worldwide due to its flexibility and standard for interconnecting multiple vendors, low operating cost and the high quality of service it provides. Plesiochronous Digital Hierarchy (PDH) on the other hand has been used before the introduction of the SDH standard and it also provides a means to transport large quantity of data via digital equipment such as radio wave systems, optic fibre and microwaves. In this paper, we shall discuss both PDH and SDH technologies and identify some of the features of both and the issues in PDH that brought about the introduction of the SDH technology.

***Keywords*** - *Networking, PDH, SDH, SONET, Telecommunication.*

## I. INTRODUCTION

In Modern telecommunication systems, the increasing demand for new services, like video and data, calls for more complicated transmission methods, higher communication speeds, and more complex network topologies. These requests, in turn, impose high design accuracy and perfect synchronization techniques of data signals [5]. The term 'Synchronization' is nowadays broadly used in telecom to encompass the methods that enable oscillators at different locations to be set to the same frequency within specified limits. With the introduction of Pulse Code Modulation (PCM) for telephony in the late 1960s which allows a single line to be used by multiple signals; using a digital time-domain multiplexing where the analog telephone signal is sampled, quantized and transmitted, network communications were being changed into digital technology and the demand for a bigger bit rate also increased. Plesiochronous Digital Hierarchy (PDH) was introduced by ITU-T G.702 [1] to cope with the increasing demand for higher bit rates; it uses a basic multiplex of 2Mbps with other stages of 8, 34 and 140Mbps. Due to the fact that PDH wasn't quite synchronous, multiplexers uses a little overhead on their high speed trunks to help cater for the differences in the data rates of streams in ports with low speed.

Due to the varying developments adopted by different networks, interconnecting gateways between networks was expensive and difficult; also PDH was not flexible which made monitoring and management more difficult to realize. Synchronous Digital Hierarchy (SDH) was developed to fix some of the limitations experienced in PDH. As more people began to use SDH, management capabilities increased because of the comprehensive monitoring and the high capability management throughout the network [4].

## II. PLESIOCHRONOUS DIGITAL HIERARCHY (PDH)

Plesiochronous Digital Hierarchy (PDH) was the standard originally for telephone networks. PDH uses time division multiplexing [3]. It was also designed to support digital voice channels running at 64kbps, was designed to use a No Store and Forward method which puts a strict restriction between the Transmitter (TX) and the Receiver (RX) and a Plesiochronous method was adapted for use which implies (nearly synchronous) [3]. PDH networks evolved, as isolated links connecting analog switching systems for Public Switched Telephone Networks.

Different standards were used in PDH which made it difficult to connect different networks. The figure below shows the different hierarchy adopted in PDH for US, Europe and Japan.

Table1. T-, E- Hierarchy [3]

| Level | US (T-) | Europe (E-) | Japan |
|---|---|---|---|
| 0 | 0.064 Mb/s | 0.064 Mb/s | 0.064 Mb/s |
| 1 | 1.544 Mb/s | 2.048 Mb/s | 1.544 Mb/s |
| 2 | 6.312 Mb/s | 8.488 Mb/s | 6.312 Mb/s |
| 3 | 44.736 Mb/s | 34.368 Mb/s | 32.064 Mb/s |
| 4 | 274.176 Mb/s | 139.264 Mb/s | 97.928 Mb/s |

The T-1 carrier system is adopted as a United States (US) standard; it uses 24-voice channels which are the results of quantization, sampling and coding using TDM framing and the PCM standard [3]. Also additional signaling channel of 1 bit is used and the T-1 speed is 1.544Mbps.

*A. Multiplexing techniques in PDH*
To move a multiple of 2Mbps data from one point to another, these data streams are multiplexed in groups of four, which is done by taking one bit from the first stream and one bit from the second stream and one bit from every other stream. In multiplexing, the transmitting multiplexer adds additional bits in order for the receiving end to decode which bits belongs to a particular 2Mbps stream of data in order to reconstruct the original data streams. The additional bits added are called "stuffing bits" or "justification bits".

*B. PDH Synchronization*
In PDH, every device has its own clock making network wide synchronization impossible. Also, errors occur during synchronization because every clock is different. The solution to preventing this error is by inserting and removing surfing bits to the frame called bit stuffing [3]. The problem of synchronization is solved by Frame Alignment Word (FAW).
If a multiplexer clock rate is higher than the tributary rate, it is called positive stuffing and this can be used for up to 140Mbps systems. On the other hand, if the multiplexer clock is lower than the tributary rate, it is called negative stuffing. When the MUX clock rate and the tributary bit rate are the same, it is called positive-negative stuffing or justification. In positive stuffing, the steps are performed:
- Data is written in a temporary buffer.
- When transmitting the data to a faster transmission channel, data is read from the buffer at a higher rate.
- Whenever the buffer is meant to be empty, a stuffing bit is normally transmitted rather than the actual data itself.
- When a stuffing bit is to be sent, a signal is sent to the receiver so that the stuffing bit can be removed at the receiving end.

In PDH, because a different frame is used both on the transmission and data layer, multiplexing and de-multiplexing operation is very complex.

*C. Limitations of PDH*
  *1) PDH is not flexible:* The difficulty involved in identifying individual channels in a higher bit stream order means that multiplexing must be performed for the high bit rate channel down through all multiplexing levels until the ideal rate is located, this requires a lot of multiplexing cost and its expense.
  *2) It is inefficient:* In PDH, it is difficult to get slower tributaries from high speed rates [3].
  *3) Lack of performance:* Since the performance of PDH systems cannot be monitored, it is difficult to provide a good performance to the system. Also there are no international agreed standards for monitoring the performance of PDH and no management channels.
  *4) PDH lacks standards:* Every manufacturer has its own standards; PDH also has different multiplexing hierarchies making it difficult to integrate interconnecting networks together.
  *5) Inefficiency in high bandwidth connections:* PDH is not ideally suited for high capacity or high bandwidth connections [3].

Other disadvantages and limitations of PDH include:
- Accessing lower tributary requires the whole system to be de-multiplexed.
- The maximum capacity for PDH is 566Mbps which is limited in bandwidth.
- Tolerance is allowed in bit rates.
- PDH allows only Point-to-Point configuration.
- PDH doesn't support Hub.

*D. Migrating to Synchronous Digital Hierarchy (SDH)*
Since the emergence of standard bodies in the 1990's, Telecom providers sparked the standardization process since the PDH system was no longer scalable to accommodate high capacity bandwidth and was not guaranteed to support traffic growth [3]. The optical technologies were gradually becoming commonly used and interoperability amongst different providers was very difficult. Synchronous Optical Network (SONET) was the American standard and Synchronous Digital Hierarchy (SDH) was proposed and the European standard.

SDH provided a vendor independent and sophisticated structure that resulted into the development of new applications, new network equipments and management flexibility than the PDH. Other services that resulted due to the evolution of SDH include:
- a) High / low speed data
- b) LAN interconnection
- c) Voice
- d) Services such as HDTV
- e) Broadband ISDN.

## III. SYNCHRONOUS DIGITAL HIERARCHY (SDH)

Unlike PDH, SDH is based on repeated hierarchy of fixed length frames that are designed to carry isochronous traffic channels. It eliminates mountains of multiplexers by allowing single stage multiplexing and de-multiplexing thereby reducing hardware complexities [1]. Some of the recommendations for the development of SDH were to define a structured multiplexing hierarchy, to define a proper protection and management mechanism, to define (optical components) physical layer requirements and to define multiplexing of different sources over SDH. The basic concept for data rates in SDH is four times the data rate for twice the cost and the table below shows the most common rate for SONET/SDH.

Table2. SONET/SDH digital hierarchy

| SONET name | SDH name | Line rate (Mbps) | Synchronous Payload Envelope rate (Mbps) | Transport Overhead rate (Mbps) |
|---|---|---|---|---|
| STS-1 | None | 51.84 | 50.112 | 1.728 |
| STS-3 | STM-1 | 155.52 | 150.336 | 5.184 |
| STS-12 | STM-4 | 622.08 | 601.344 | 20.736 |
| STS-48 | STM-16 | 2,488.32 | 2,405.376 | 84.672 |
| STS-192 | STM-64 | 9,953.28 | 9,621.504 | 331.776 |
| STS-768 | STM-256 | 39,813.12 | 38,486.016 | 1,327.104 |

Designed to optimize Time Division Multiplexing (TDM) SONET/SDH is very reliable and contains a built in system to provide 99.99% uptime [2]. Some of the broad features of SONET/SDH include:
- SONET/SDH uses Time Division Multiplexing.
- It encompasses optical and electrical specifications.
- It uses octet multiplexing.
- Also uses an extremely precise timing.
- SDH provides support for operation maintenance and administration.
- Improvement over T-carriers identifies sub streams, international connectivity and enhanced control and administrative functions.
- SONET/SDH also fits in the physical layer.

### A. SDH Network elements
The different network elements in SDH include:

*1) Synchronous multiplexer:* The synchronous multiplexer performs both the live line transmitting functions and multiplexing, it replaces Plesiochronous multiplexers and line transmitting equipments. There are two types of synchronous multiplexers;
- *Terminal Multiplexers (TM):* These multiplexers accept a number of tributary signals and multiplex them into appropriate aggregate signals.
- *Add and Drop Multiplexers (ADM):* ADM allows it to be possible to "ADD" channels or "DROP" channels from "THROUGH CHANNELS". It is SDH building block for local access to synchronous network [4].

*2) Synchronous Digital Cross Connect:* The cross connect equipment acts as a switch that can pick out one or more lower order channels without the need for a transmission channel.

*3) Regenerators:* This is a device that regenerates the signals. The major use of regenerators is for long distance data transfer of more than 50km, termination is performed and the optical signal is regenerated [4].

### B. SDH Frame Structure
The SDH frame structure is based on synchronous byte-wise multiplexing of several building blocks. Such synchronous multiplexing elements are structured fixed-size sets of bytes, which are byte-interleaved or mapped one-into-the-other to eventually form STM-N frames. The STM-1 frame is the basic transmission format for SDH. The frame lasts for 125µs, equivalent to 0.125 kHz.

### C. SDH Virtual Container Structure
Virtual containers (VCs) are the basic building block, which maps a payload that can be any PDH signal as well as other lower-order synchronous multiplexing elements [5]. VCs are individually and independently accessible within SDH frames through pointer information directly associated with them by multiplexing. These overhead bytes are added whenever the layer is introduced and removed when the layer is terminated.

*D. Structure of SDH Overheads*
The SDH Overheads support: monitoring, messaging, labeling, and switching control. In each layer, specific bytes are allocated per frame, or per multi-frame; Overheads allow for monitoring of both ends from one end, for sector management (transit traffic), and central management via Data Communications Channel (DCC).

*E. SDH Layers*
The SDH layer consists of four sub-layers, which are path, regenerator section, photonic layer and multiplex section. The SDH framing structure defines overheads operating at these layers to estimate error rate, communicate alarm condition and provide maintenance support.

*F. SDH Network Topology*
  *1) Point-Point Link:* Based on PDH systems which provide point – point connections, SDH will replace these systems with STM-4 line systems [4]. In this system, regenerators may be used to avoid transmission issues, no routing or de-multiplexing is done along the path [3].
  *2) Ring Topology:* This is the most used topology, in this topology, two or four fibers can be used and an ADM at each node [3]. The ring network is a route back to itself that facilitates the development of protocols that can detect if there is a failure in the fibre and re-establish connection back quickly.
  *3) Star Topology:* The traffic here passes through a central hub where the hub is a Synchronous Digital Cross Connect [4].
  *4) Linear Bus Topology:* The Linear bus topology has great flexibility and is used when there is a need for protection [4].

*G. Advantages of SDH*
Comparing SDH to PDH, the transmission rates of SDH can go up to 10Gbps, it is easier to extract and insert low-bit rate channels to high bit streams. SDH systems include auto backup and restore/repair mechanisms in case of failure, and a failure in a link or a network element does not amount to the failure of the entire network. Other advantages of SDH include:
- A more simplified multiplexing and de-multiplexing technique.
- Synchronous networking and SDH supports multipoint networking [4].
- Capability of transporting existing PDH signals.
- Easy growth to higher bit rates which enhances the administration and maintenance process.
- It's capable of transporting broadband signals.
- It's multivendor and supports different operators [4].
- It provides network transport services on LAN such as video conferencing, and interactive multimedia [4].
- Optical fibre bandwidth can be increased without limit in SDH [4].
- Switching protection to traffic is offered by rings.
- SDH allows quick recovery from failure.

*H. Future of SDH*
Most fibre optic public network transmission systems now use SDH, it is expected to dominate transmission for years to come and since SDH provides flexible means for this via interoperability and high capacity, more and more development and services with high bandwidth consumption will be provided to customers. Also in the future, increased Internet and intranet uses of applications such as online gaming, high-volume remote data backup and integrated voice and video over fiber to the home (FTTH) will be paramount and easy to implement due to the introduction of SDH.

## IV. CONCLUSION
The optical networks are becoming more and more important with increasing demand for the availability of high-speed networks. By combining highly accurate network synchronization systems with advanced optical network technology. High-speed transport systems like SDH Network Elements can guarantee the high performance levels that users will demand from current and future telecommunications systems. Although SDH provides a means for this, the system is a lot more complex than explained in this paper. SDH solves most of the problems encountered in PDH as faster communication networks where developed such as the optical fibre networks. With the universal standard provided by SDH, more independent manufacturers can begin to innovate more on providing higher bandwidth capacity services to customers by increasing optical fibre bandwidth without any extra requirement with SDH.

## V. REFERENCES

[1] A. V. Palle, "SDH (Synchronous Digital Hierarchy) & Its Architecture," Int. J. Sci. Res. Dev., vol. 1, no. 7, pp. 1–5, 2013.
[2] B. W. Habisreitinger, "Next Generation SONET / SDH – Technologies and Applications," JDSU, 2006.
[3] F. Neri and G. Reti, "SONET-SDH," Politec. di Torino, Dep. Electr. Eng., 2007.
[4] N. Jyothirmai, R. M. Valli, and A. R. Krishna, "SDH and Its Future Trends," Int. J. Innov. Technol. Explor. Eng., vol. 1, no. 6, pp. 74–78, 2012.
[5] S. Manke, K. Khare and S. Sapre, "100Mbps Ethernet data transmission over SDH networks using Cross Virtual Concatenation", pp. 1--6, 2008.